\documentclass[aps,superscriptaddress,twocolumn,amssymb,longbibliography]{revtex4-1}

\usepackage{amsmath}
\usepackage{amssymb}
\usepackage{graphicx}
\usepackage{braket}
\usepackage{hyperref}

\begin{document}
\title{Contextuality without nonlocality in a superconducting quantum system}

\author{Markus Jerger}
\affiliation{ARC Centre of Excellence for Engineered Quantum Systems, The University of Queensland, St Lucia QLD 4072
	Australia}
\author{Yarema Reshitnyk}
\affiliation{School of Mathematics and Physics, University of Queensland, Brisbane, Queensland 4072, Australia}

\author{Markus Oppliger}
\affiliation{Department of Physics, ETH Zurich, CH-8093 Zurich, Switzerland}

\author{Anton Poto\v{c}nik}
\affiliation{Department of Physics, ETH Zurich, CH-8093 Zurich, Switzerland}

\author{Mintu Mondal}
\affiliation{Department of Physics, ETH Zurich, CH-8093 Zurich, Switzerland}

\author{Andreas Wallraff}
\affiliation{Department of Physics, ETH Zurich, CH-8093 Zurich, Switzerland}

\author{Kenneth Goodenough}
\affiliation{QuTech, Delft University of Technology, Lorentzweg 1, 2611 CJ Delft, Netherlands}

\author{Stephanie Wehner}
\affiliation{QuTech, Delft University of Technology, Lorentzweg 1, 2611 CJ Delft, Netherlands}

\author{Kristinn Juliusson}
\affiliation{Quantronics, SPEC, IRAMIS, DSM, CEA Saclay, Gif-sur-Yvette, France}

\author{Nathan K.\ Langford}
\affiliation{QuTech, Delft University of Technology, Lorentzweg 1, 2611 CJ Delft, Netherlands}
\affiliation{Kavli Institute of Nanoscience, Delft University of Technology, P.O. Box 5046, 2600 GA Delft, Netherlands}

\author{Arkady Fedorov}
\affiliation{ARC Centre of Excellence for Engineered Quantum Systems, The University of Queensland, St Lucia QLD 4072
	Australia}
\affiliation{School of Mathematics and Physics, University of Queensland, Brisbane, Queensland 4072, Australia}

\begin{abstract}
Classical realism demands that system properties exist independently of whether they are measured, while noncontextuality demands that the results of measurements do not depend on what other measurements are performed in conjunction with them. The Bell-Kochen-Specker theorem states that noncontextual realism cannot reproduce the measurement statistics of a single three-level quantum system (qutrit). Noncontextual realistic models may thus be tested using a single qutrit without relying on the notion of quantum entanglement in contrast to Bell inequality tests. It is challenging to refute such models experimentally, since imperfections may introduce loopholes that enable a realist interpretation. Here we use a superconducting qutrit with deterministic, binary-outcome readouts to violate a noncontextuality inequality while addressing the detection, individual-existence and compatibility loopholes. 
This evidence of state-dependent contextuality also demonstrates the fitness of superconducting quantum circuits for fault-tolerant quantum computation in surface-code architectures, currently the most promising route to scalable quantum computing.

\end{abstract}

\date{\today}

\maketitle

Realistic models of nature aim to describe the predictions of quantum mechanics using underlying hidden variables (HVs),
which determine the properties of the system ahead of time.
The best known examples are local HV theories, which seek to explain the predictions of quantum entanglement under the combined assumptions of realism and locality~\cite{Bell1964}.
The divide between quantum and classical physics, however, runs deeper than the feature of entanglement.
The Bell-Kochen-Specker theorem~\cite{Bell1966, Kochen1967} considers nonconextual HV models which are defined without reference to locality.
The Bell-Kochen-Specker theorem shows that, already for qutrit systems, it is not possible to define such a model that is consistent with quantum theory.

While the original theorem is difficult to test, the discovery of noncontextuality inequalities~\cite{Klyachko2008, Cabello2008, Araujo2013, Yu2012} makes tests of noncontextual models accessible experimentally even in the presence of imperfections.
Noncontextuality tests have been carried out in a range of different physical systems and dimensionalities, including neutrons~\cite{Bartosik2009}, trapped ions~\cite{Kirchmair2009,Zhang2013}, single photons~\cite{Lapkiewicz2011,Marques2014b,Mazurek2015} and spins of nitrogen-vacancy (NV) centers in diamond~\cite{George2013,Kong2016}, but all of these experimental tests introduced additional loopholes.
As in tests of local realism, insufficient detector efficiencies lead to the detection loophole. 
Here, ignoring undetected events introduces a selection bias which can be exploited to find a consistent HV explanation.
The individual-existence and compatibility loopholes are important for noncontextuality tests, which require the comparison of multiple outcomes in a measurement context~\cite{Kochen1967, Klyachko2008, Peres2002}.
If measurements are performed jointly~\cite{Arias2015}, it is not always possible to establish a meaningful operational definition of an individual measurement.
It is therefore critical to obtain individual measurement outcomes for each measurement, for example by making measurements sequentially~\cite{Larsson2014}.
The compatibility loophole arises when imperfections cause sequential measurements to be imperfectly commuting~\cite{Kirchmair2009, Guhne2010, Szangolies2013}.
The compatibility loophole has been addressed in both photonic~\cite{Marques2014b} and trapped-ion~\cite{Kirchmair2009} experiments, but the detection and individual-existence loopholes have only been addressed using high-efficiency read-out in trapped ions~\cite{Kirchmair2009,Zhang2013}.
The three loopholes have only been addressed simultaneously in a two-qubit scenario~\cite{Kirchmair2009}, where it remains possible to construct explanations involving quantum entanglement.

In this experiment, we realize the Klyachko-Can-Binicio\u{g}lu-Shumovsky (KCBS) state-dependent noncontextuality test~\cite{Klyachko2008} with a tunable superconducting qutrit.
By engineering deterministic, binary-outcome readouts, we violate a noncontextuality inequality while addressing the detection, individual-existence and compatibility loopholes in a singe experiment without entanglement. 
This evidence of state-dependent contextuality in superconducting quantum circuits demonstrates their suitability for fault-tolerant quantum computation using magic state distillation~\cite{Howard2014}.

\section{Results}

\begin{figure}[htbp]
	\begin{center}
		\includegraphics[viewport = 0 0 150 150,scale=1]{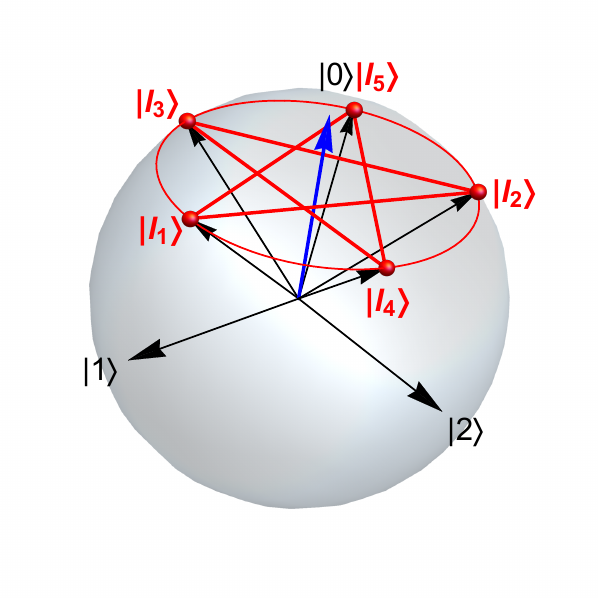}
	\end{center}
	\caption{KCBS pentagram: The qutrit eigenstates are $|i\rangle$ with $i=0,1,2$. One can construct five qutrit states $|l_i\rangle$ corresponding to five dichotomic observables $A_i = 2|l_i\rangle\langle l_i| - 1$.  States connected by edges of the pentagram are orthogonal, assuming compatibility of the associated observables.  Each pair of compatible measurements forms a context, and each observable is included in two different contexts. The states of the pentagram are chosen to provide maximum contradiction with noncontextual HV models.
	}
	\label{fig:1}
\end{figure}

{\bf The Klyachko-Can-Binicio\u{g}lu-Shumovsky test.} The Klyachko-Can-Binicio\u{g}lu-Shumovsky (KCBS) state-dependent noncontextuality test~\cite{Klyachko2008} uses five different observables $A_i$ ($i = 1,2,...,5$) with binary outcomes $\pm 1$. The test involves measuring the five pairs of observables, called measurement contexts, $\{A_1,A_2\}$, $\{A_2,A_3\}$, $\{A_3,A_4\}$, $\{A_4,A_5\}$ and $\{A_5,A_1\}$, chosen such that each observable is measured in two different contexts.
Noncontextual HV models predict that the total observable correlations for outcome pairs are bounded by~\cite{Klyachko2008}
\begin{equation}\label{KCBS}
\langle A_1 A_2\rangle + \langle A_2 A_3\rangle + \langle A_3 A_4\rangle + \langle A_4 A_5\rangle + \langle A_5 A_1\rangle \geq -3.
\end{equation}
This inequality can be violated in quantum mechanics. Here, we consider a qutrit system, with five dichotomic observables $A_i = 2 |l_i\rangle\langle l_i| -1$ corresponding to states represented by vertices of the pentagram shown in Fig.~\ref{fig:1}.
Each observable can be described by a pair of projectors $\{|l_i\rangle\langle l_i|, I-|l_i\rangle\langle l_i|\}$ associated with outcomes $\{+1,- 1\}$.
The states connected by edges of the pentagram are orthogonal, ensuring that the corresponding observables, $A_i$ and $A_{i+1}$, (and measurement operators) commute, making them compatible observables.
Quantum mechanics predicts that the left side of (\ref{KCBS}) evaluates to $5 - 4\sqrt{5} \simeq -3.944$ for a qutrit in the ground state, $|0\rangle$.
This is the maximum quantum violation of inequality (\ref{KCBS})~\cite{Araujo2013}.

\begin{figure*}[htb]
	\begin{center}
        \includegraphics[width=0.7\textwidth]{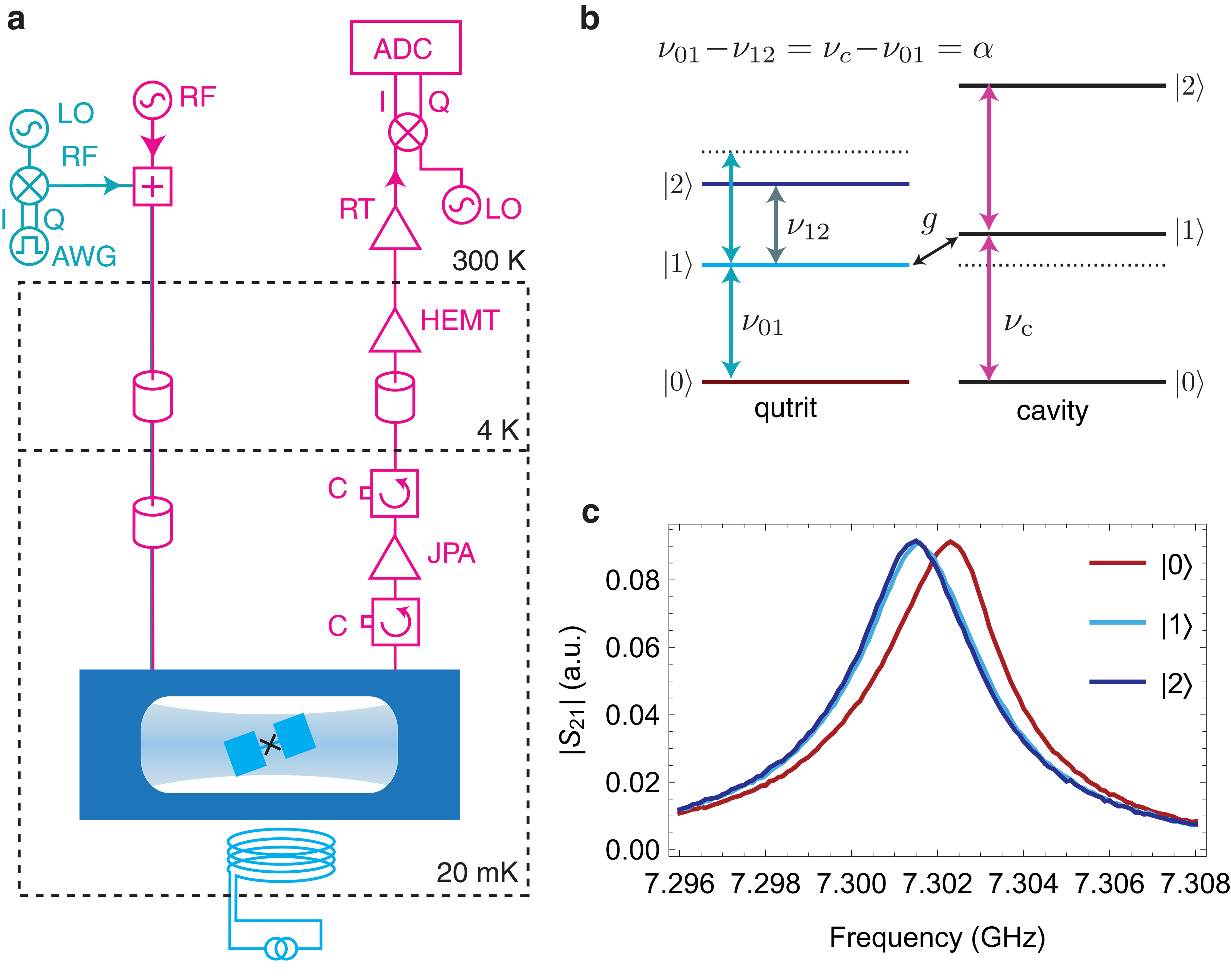}
	\end{center}
	\caption{System and measurement setup:
		a. Simplified diagram of the measurement setup (see Methods for details). 
		b. The energy level diagram of a qutrit coupled to a microwave cavity when the dispersive shifts of the cavity frequency are identical for the first and second excited states of the qutrit. The scheme realizes the binary outcome projective measurement of the qutrit on its ground state $M_{|0\rangle}$. 
		c. Transmission through the readout cavity with the qutrit in different basis states. The qutrit was first prepared in different basis states followed a square microwave pulse with a frequency close to the resonant frequency of the cavity for several microseconds. The plot indicates the normalized amplitude of measured transmitted signal integrated over $2\,\mathrm{\mu s}$. The dispersive shifts for $\ket{1}$ and $\ket{2}$ are close to identical, not allowing the measurement to distinguish between the two states. }
	\label{fig:2}
\end{figure*}

{\bf Superconducting qutrit.} We encode a qutrit into a transmon-type multilevel quantum circuit~\cite{Koch2007} incorporated into a 3D microwave copper cavity (Fig.~\ref{fig:2}a,b). 
The three lowest energy eigenstates of the weakly anharmonic transmon form the qutrit's logical states, with allowed transition frequencies of $\nu^\mathrm{max}_{01} = 6.939\,\mathrm{GHz}$ between the ground and first excited states and $\nu^\mathrm{max}_{12} = 6.623\,\mathrm{GHz}$ between the first and second excited states, corresponding to an anharmonicity of  $\alpha \equiv \nu_{12} - \nu_{01} = - 314\,\mathrm{MHz}$.
The qutrit is dispersively coupled with strength $g = 17.9\,\mathrm{MHz}$ to the cavity's fundamental mode (with bare frequency $\nu_\mathrm{c} = 7.3014\,\mathrm{GHz}$ and linewidth $2.4\,\mathrm{MHz}$).
We detect the state of our transmon qutrit in the usual way via the state-dependent frequency shift of the cavity, by measuring the amplitude and phase of a probe signal transmitted through the cavity.
This signal is then amplified by a Josephson parametric amplifier~\cite{Eichler2014}, a cryogenic high electron mobility transistor amplifier and a chain of room-temperature amplifiers.
The high-fidelity single-shot detection enabled by the parametric amplifier ensured that each experimental trial produced a definite outcome, thus closing the detection loophole.

\begin{figure}[htbp]
	\begin{center}
		\includegraphics[width=\columnwidth]{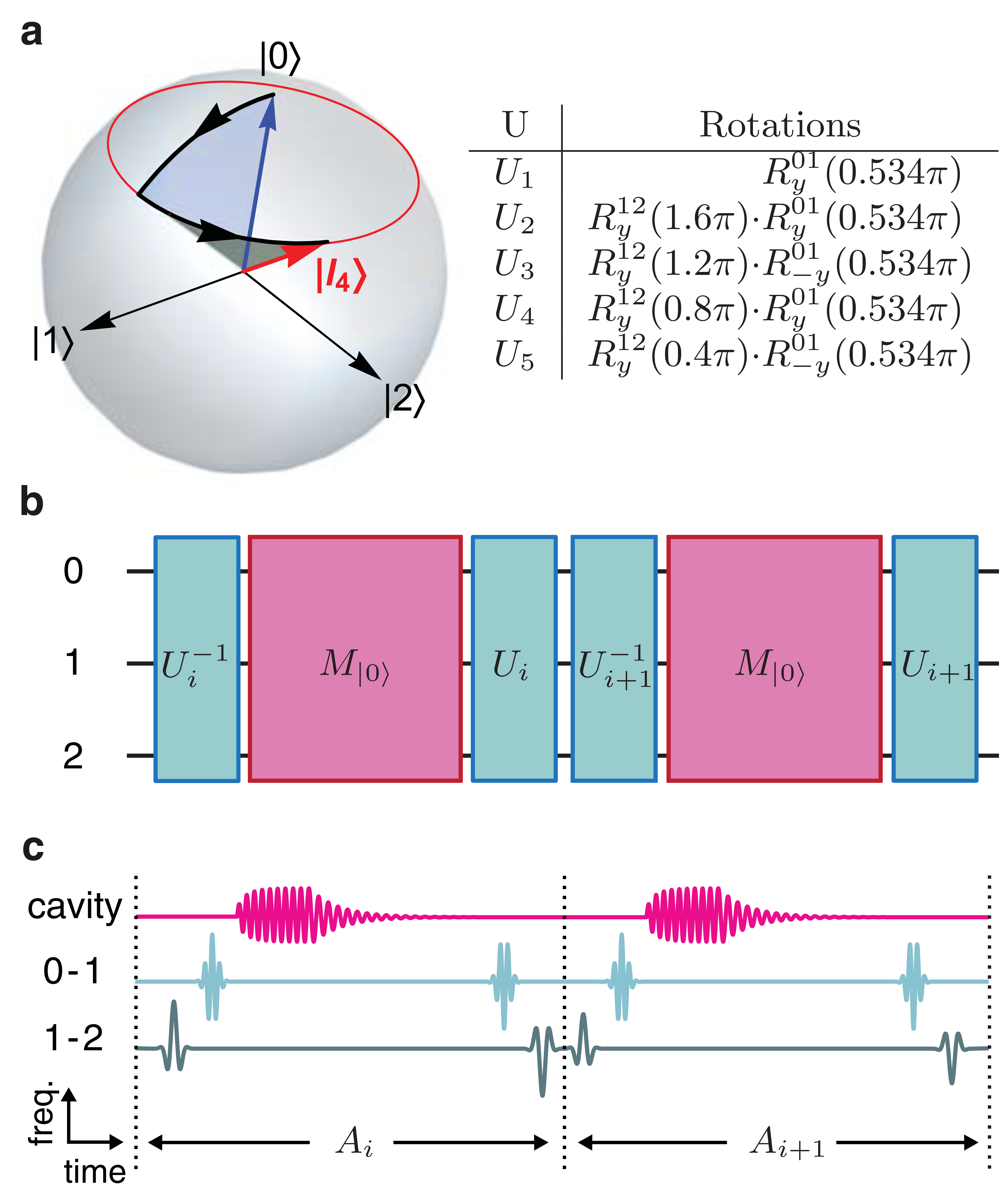}
		\caption{Measurement potocol:
                         a. Unitary transformations of the qutrit ground state to the KCBS states. Each $U_i$ can be decomposed into one or two rotations $R_{\hat n}^{i,i+1}(\phi)$, where $\phi$ is a rotation of angle about the axis $\hat n$ in the qutrit subspace spanned by $\{|i\rangle, |i+1\rangle\}$. The rightmost pulse in a product is applied first in time. The trajectory of the state under transformation $U_4$ is shown as an example.
                         b. The measurement protocol includes two sequential projecting measurements $M_{|0\rangle}$ onto the ground state with unitary transformations before and after each measurement. The unitaries rotate the measurement axis into one of the states of the KCBS pentagram.
                         c. The actual experimental sequence for each pair of measurements. Measurement of the $M_{|0\rangle}$ observable is implemented with a cavity probe signal and the qutrit rotations $R_{\hat n}^{i,i+1}(\phi)$ are constructed with microwave pulses applied at the qutrit transition frequencies.
		}
		\label{fig:3}
	\end{center}
\end{figure}

{\bf Binary-outcome readout.} In this experiment, we close the individual-existence loophole by performing efficient sequential measurements with classical, binary outcomes.
Critical to this is our ability to implement partially projective dichotomic measurements on the qutrit system.
For our transmon system, the state-dependent cavity frequency shift is~\cite{Koch2007}
\begin{equation}
s_j = -\chi_j + \chi_{j-1}, \qquad \chi_{j\geq 0} \approx \frac{(j+1) g^2}{\nu_{j,j+1} - \nu_\mathrm{c}}, \quad \quad \chi_{-1} = 0.
\end{equation}
Ordinarily, this gives distinguishable measurement responses for all states of the qutrit~\cite{Bianchetti2010}, resulting in a fully projective measurement. By choosing the qutrit detuning $\nu_{01} - \nu_\mathrm{c}\simeq \alpha $ (see Fig.~\ref{fig:2}b), we match two of the dispersive shifts, $s_1$ and $s_2$ (Fig.~\ref{fig:2}c), making the corresponding measurement responses indistinguishable. 
Probing the cavity therefore quickly erases coherences between the ground state and the excited states, but leaves the coherence between the first and second excited states intact, realizing a dichotomic measurement along $\ket{0}$, associated with the observable $M_{|0\rangle} = 2\ket{0}\bra{0}-1$ with projectors $\{|0\rangle\langle0|, I -|0\rangle\langle0|\}$.
For detailed information on the effect of the readout pulse on the state of a qutrit state for different detunings, including a theoretical model, experimental verification and calibration procedures see Ref.~\cite{Jerger2015}). 
Probing the cavity for $350\,\mathrm{ns}$, we reach a single-shot contrast of $\approx 96\%$ between $\ket{0}$ and $\ket{1}, \ket{2}$ limited primarily by thermal excitation and decay of the qutrit state during the readouts. 
To generate measurements in arbitrary directions $\ket{l_i}$ from $M_{|0\rangle}$, we apply unitary rotations before and after measurement (see Fig.~\ref{fig:3}a).
Each measurement procedure starts by sending microwave pulses to the qutrit to rotate the desired measurement basis (defined by one of the KCBS states $\ket{l_i}$; see Fig.~\ref{fig:3}b,c) onto the ground-state readout basis.
After a readout pulse and delay of 475\,ns for cavity ring-down, further microwave pulses return the qutrit to its initial reference frame, necessary to allow subsequent measurements to be implemented independently.

{\bf Testing compatibility.} Preserving coherence in the subspace orthogonal to the measurement direction is also crucial for ensuring the compatibility of context-independent sequential measurements.
Since noncontextuality tests aim to falsify noncontextuality using the assumptions of noncontextual realism, which contain no notion of compatibility, it is important to ask why test protocols only consider compatible measurements.
It is well established that individual outcome probabilities for incompatible observables will depend on the order in which they are measured, but this overt contextuality does not reveal any further insight into the nature of reality.
However, restricting attention to compatible measurements allows a study of whether context dependence still remains when this overt contextuality is absent.
In practice, experimental imperfections make the actual measurement procedures only approximately compatible. This loophole can be addressed by an extended KCBS inequality~\cite{Guhne2010}
\begin{multline}\label{KCBS2}
\langle A_1 A_2\rangle + \langle A_3 A_2\rangle + \langle A_3 A_4\rangle + \langle A_5 A_4\rangle + \langle A_5 A_1\rangle 
\geq \\
-3 - \left(\varepsilon_{12} + \varepsilon_{32} + \varepsilon_{34} + \varepsilon_{54} + \varepsilon_{51}\right).
\end{multline}
Here, the order of the observables in the two-outcome correlations $\langle A_i A_{j}\rangle$ corresponds to the timing order for two corresponding sequential measurements, and $\varepsilon_{ij}$ are the operational bounds for the incompatibility of these measurement procedures. A bound on incompatibility~\cite{Guhne2010}:
\begin{equation}\label{eq:incomp}
\varepsilon_{ij} = |\langle A_j| A_j A_i\rangle - \langle A_j|A_i A_j\rangle|,
\end{equation}
with $A_j$ measured before/after $A_i$ can be established separately (see Supplementary Information).

{\bf Protocol and measurements of correlations.} In the final protocol, we measure the five combinations $\braket{A_1 A_2}$, $\braket{A_2 A_3}$, $\braket{A_3 A_4}$, $\braket{A_4 A_5}$, $\braket{A_5 A_1}$ and their reverse-order variants, followed by calibration blocks to detect phase drifts of the cavity signal.
As the qutrit is operated in a dilution refrigerator at 20\,mK, its thermal state is close to the ground state.
To avoid residual thermal population, we begin each experimental trial by applying an initialisation readout tone to the cavity to project and post-select the desired ground state, rejecting approximately 10\% of all data points.
A further delay of 565\,ns allows the cavity to ring down before the measurement sequence begins.
The whole sequence is repeated $2^{21}$ times, triggered every $100\,\mathrm{\mu s}$.
For each observable $A_i$, the same procedure (set of pulses) was used to implement the measurement independent of measurement context, with the cavity transmission signal detected, integrated and discriminated to assign a dichotomic outcome $\pm1$. 
The outcomes were recorded to the hard drive and were later used to calculate expectation values $\langle A_i \rangle$ and correlations $\langle A_iA_j\rangle$.
The results used to test inequality~(\ref{KCBS2}) (and its reverse-order counterpart) are presented in Table~\ref{Table: results}).
For all pairs, the first measurement yields expectation values very close to the ideal value of $0.105(6)$, with the second measurement consistently displaced due to decoherence of the qutrit during the sequence.
We find a sum of correlations of -3.489(1) and the threshold including incompatibility bounds of -3.352(2).

\begin{table*}[hbt]
	\small\center
	\begin{tabular}{|c|cc|cc|cc|}
		\hline
		(i,j) & \multicolumn{2}{|c|}{$\braket{A_i A_j}$}&  $\braket{A_i}$ & $\braket{A_j}$ & \multicolumn{2}{|c|}{$\varepsilon_{ij}$}\\
		\hline
		(1,2) & -0.6947(5) & & 0.0744(7) &  0.1475(7) &  0.073(1) &\\
		(2,1) & &-0.7009(5) &  0.0741(7) &  0.1530(7) & & 0.079(1) \\
		(2,3) & &-0.7080(5) &  0.0748(7) &  0.1470(7) & & 0.072(1) \\
		(3,2) & -0.7001(5) & & 0.0808(7) &  0.1488(7) &  0.068(1) &\\
		(3,4) & -0.6907(5) & & 0.0820(7) &  0.1551(7) &  0.073(1) &\\
		(4,3) & &-0.6996(6) &  0.0784(7) &  0.1511(7) & & 0.073(1) \\
		(4,5) & &-0.6992(5) &  0.0781(7) &  0.1500(7) & & 0.072(1) \\
		(5,4) & -0.7051(5) & & 0.0768(7) &  0.1477(7) &  0.071(1) &\\
		(5,1) & -0.6986(5) & & 0.0779(7) &  0.1452(7) &  0.067(1) &\\
		(1,5) & &-0.7052(5) &  0.0753(7) &  0.1469(7) & & 0.072(1) \\
		\hline
		\hline
		$\sum$  & -3.489(1) & &          &          &  0.352(2)&\\
		& &-3.513(1) &          &          &&  0.367(2)\\
		\hline

	\end{tabular}
	\caption{Violation of the KCBS inequality. 
		Correlations $\braket{A_i A_j}$ contribute to the left side of Eq.~($\ref{KCBS2}$). We also provide $\braket{A_i A_j}$ for the equation with the reversed order of measurements. Single expectation values $\braket{A_i}$ and $\braket{A_j}$ are used to evaluate bounds $\varepsilon_{ij}$ on compatibility contributing to the right side of Eq.~($\ref{KCBS2}$). 
		Inequality $\sum \braket{A_i A_j} \geq -3 - \sum \varepsilon_{ij}$ is experimentally violated for forward and reversed orders by more than 53 and 56 standard deviations, respectively.}\label{Table: results}
\end{table*}

{\bf $P$-value calculation.} Using the standard analysis of such inequalities, we violate the KCBS noncontextuality inequality Eq.~(\ref{KCBS2}) by more than 53 standard deviations.
Inspired by the extended inequality derived in Ref.~\cite{Guhne2010}, the compatibility loophole was treated by  formalizing the problem as a hypothesis test without any assumptions on compatibility and bounding incompatibility of the measurements in a separate hypothesis test~\cite{Elkouss2015}.
The null hypothesis that the experiment is described by a noncontextual HV model with compatibility $\epsilon \lesssim 0.0413$ (see Methods)
is rejected with a $P$-value less than $2.96 \times 10^{-575}$. A separate test of the compatibility condition rejects the hypothesis that the observables are more incompatible with a $P$-value less than $4.1 \times 10^{-4}$. Our analysis requires only the assumption that the devices perform the same in every single trial and the no-memory assumption without any
additional assumptions on compatibility of the measurements or on the measurement contrast.


\section{Discussions}
Our results strongly contradict the predictions of noncontextual hidden variable models, closing two common loopholes: the detection loophole, via high-fidelity, deterministic single-shot readout and the individual-existence loophole~\cite{Larsson2014}, using separate, sequential measurements. The compatibility loophole was treated by violating an extended inequality~\cite{Guhne2010} and, independently, by formulating the problem in the form of a hypothesis test without any assumptions on compatibility and bounding the incompatibility of the measurements in a separate hypothesis test. 

As a key ingredient in addressing these loopholes, we implemented sequential dichotomic qutrit measurements which project out one target state without disturbing the information stored in the remaining two-dimensional subspace. This allows a classical result from the first measurement to be obtained before implementing the setting to be used for the second measurement. Our results demonstrate that quantum mechanics departs from predictions of noncontextual realism, without reliance on nonlocality or entanglement correlations, and provide evidence of the contextuality resource in superconducting circuits. While we used the simpler state-dependent inequality for demonstration of contextual nature of the superconducting circuits, the state-independent test will be the straightforward extension of our experiment.

One key point which differentiates our noncontextuality analysis for an indivisible system from a similar analysis for Bell inequalities with locality constraints, is that it is difficult to avoid the need for additional i.i.d. and no memory assumptions for measurements on a single system. Since quantum contextuality can be simulated by a classical system with memory~\cite{Kleinmann2011}, these loopholes will most likely remain for any Kochen-Specker tests without nonlocality. In another case, for the finite-precision loophole~\cite{Meyer1999, Kent1999}, debate continues about whether this loophole can be closed in principle~\cite{Simon2001, Cabello2011, Winter2014}. It
remains an important open challenge to identify a clear, general prescription for how to implement a noncontextuality test with minimal assumptions.

\section{Methods}

\subsection{Sample and cavity.}

The qutrit was fabricated on an intrinsic Si substrate in a single step of electron beam lithography followed by shadow evaporation of two Al layers with an oxidation step between the depositions. The design of the circuit is identical to the one in Ref.~\cite{Paik2010} and consists of two sub-millimetre size capacitor plates connected via a line interrupted by a d.c. superconducting quantum
interference device (SQUID), playing the role of a magnetically tunable Josephson junction. Magnetic flux supplied by a superconducting coil attached to the copper cavity is used to control the transition frequencies of the qutrit. The qutrit has maximum transition frequencies of $\nu^\mathrm{max}_{01} = 6.950\,\mathrm{GHz}$ between the ground and first excited states, $\nu^\mathrm{max}_{12} = 6.635\,\mathrm{GHz}$ between the first and second excited states, corresponding to a anharmonicity of  $\alpha = 314\,\mathrm{MHz}$ and charging energy of $E_C/h = 288\,\mathrm{MHz}$ as shown in Fig.~\ref{fig:2}b. At the working point of the qutrit, selected by the magnetic field bias, we measured decay and coherence times of
$T_{1,1} = 17.4\,\mathrm{\mu s}$,
$T_{1,2} = 6.2\,\mathrm{\mu s}$,
$T_{1,2 \rightarrow 1} = 18.1\,\mathrm{\mu s}$,
$T_{1,2 \rightarrow 0} = 9.5\,\mathrm{\mu s}$,
$T^*_{2,01} = 6.6\,\mathrm{\mu s}$ and
$T^*_{2,12} = 4.6\,\mathrm{\mu s}$.

The qutrit was incorporated into a 3D microwave copper cavity attached to the cold stage of a dilution cryostat (see Fig.~\ref{fig:2}a). The  cavity was coupled asymmetrically to the input and output microwave ports with corresponding external quality factors of $Q_{\rm in} \simeq 80\,000$ and $Q_{\rm out} = 4\,200$ for transmission measurements with the internal quality factor of the cavity was measured in the separate runs as $Q \sim 10\,000$ at mK temperatures.

To measure transmission, a signal from a microwave generator (RF) was applied to the input port of the cavity. Microwaves transmitted through the cavity were amplified by a Josephson parametric amplifier (JPA), high-electron-mobility transistor (HEMT) amplifier at 4~K and a chain of room-temperature (RT) amplifiers. The sample at 20~mK was isolated from the higher-temperature fridge stages by three circulators (C) in series. The amplified transmission signal was down-converted to an intermediate frequency of 25~MHz in an IQ mixer driven by a dedicated LO, and digitized by an analogue-to-digital converter (ADC) for data analysis.




\subsection{Readout}
To implement single-shot readout, we used a Josephson parametric dimer amplifier (JPDA)~\cite{Eichler2014} as a preamplifier of the signal. The JPDA consists of two coupled non-linear resonators and can be operated in the non-degenerate mode if a pump tone frequency is set between resonance frequencies of the resonators. In our experiment the pump tone was set at $7.058\,\mathrm{GHz}$ providing amplification of $25\,\mathrm{dB}$ gain and $12.5\,\mathrm{MHz}$ bandwidth centered at the readout frequency $\nu_\mathrm{c}$. Two circulators installed between the readout cavity and JPDA, combined with the readout cavity itself, eliminated any effect of the pump tone on the qutrit.

\subsection{Hypothesis test}
Experimental tests of HV models can be formulated as a hypothesis test, where the null hypothesis (to be rejected) is that the measurement statistics can be modelled using 
HVs~\cite{Elkouss2015}. To this end, the experiment is recast as a set of trials of a game which can be won with a maximum probability of $\beta_{\rm win}$ if the experiment were 
governed by a specific non-contextual HV model. Specifically, we tests an i.i.d. model (the devices behave the same in each trial) in which the compatibility
of measurements obeys a guaranteed limit as in~\eqref{eq:incomp} (see Supplementary Information). This limit is then tested separately.
To this end, it is convenient to phrase the compatibility condition of~\eqref{eq:incomp} in terms of probabilities instead of expectation values
as $|\Pr(A_j =a_j|\#1=j) - \Pr(A_j =a_j|\#1=i,\#2=j)|\leq \epsilon_{(i,j)}$, where we use $\#1$ and $\#2$ 
to indicate the order in which we make the measurements $A_j$ labeled $i$ and $j$, and $a_j$ denotes the outcome of measurement $j$. 
An $\epsilon$-incompatible model assumes that
\begin{align}\label{eq:condition}
\left\|\frac{1}{5}\sum_{(i,j)} \epsilon_{(i,j)}\right\| \leq \epsilon\ . 
\end{align}
For the KCBS inequality, a trial is won if the two outcomes of a context are not equal. The total number of wins is recorded over the whole experimental run of $n$ trials.
The $P$-value is then the probability that the game could have been won at least that many times given a non-contextual hidden variable model with incompatibility $\epsilon$.
In this experiment, we recorded 3912769 wins out of 4603450 trials, which implies that the $P\text{-value} \leq 2.96 \times 10^{-575}$. 
A second, parallel hypothesis test is formulated to test the incompatibility bound of~\eqref{eq:condition}.

\subsection{Data availability}The measurement data that support the findings of this study are available in UQ eSpace with the identifier(s) [data DOI(s) e.g. “http://dx.doi.org/10.14264/uql.2016.207”]



\section{Acknowledgements} 
We thank Pascal Macha, Andr\'es Rosario Hamann and Kirill Shulga for help at the early stage of the experiment. We also thank Fabio Costa and Clemens M\"uller for useful discussions. M.J., A.F. were supported by the Australian Research Council Centre of Excellence CE110001013. Y.R. was supported by the Discovery Project DP150101033. AF was supported in part by the ARC Future Fellowship FT140100338. KJ was supported by the CCQED network. KG and SW are supported by STW, and an NWO VIDI Grant.

\section{Author contributions}
AF, MJ and NKL designed the experiment. MJ and YR performed the experiment. MO, AP, MM and AW designed and fabricated the JPA. KJ designed and fabricated the qutrit. MJ carried out data analysis to calculate expectation values, correlations and standard deviations. KG and SW formulated the results in the form of hypothesis test and calculated $P$-values.  AF, MJ, NKL, SW, KG wrote the manuscript. AW, YR commented on the manuscript. AF supervised the project.

\section{Additional information}
{\bf Supplementary information} accompanies this paper.

{\bf Competing financial interests:} The authors declare that they have no competing financial interests.

\section{Correspondence} Correspondence and requests for materials should be addressed to A.F.~(email: a.fedorov@uq.edu.au).


\begin{thebibliography}{31}%

\bibitem{Bell1964}
Bell, J.~S.~ On the Einstein-Podolsky-Rosen paradox. {\it  Physics}  {\bf 1,} 195-200 (1964).

\bibitem{Bell1966}
Bell, J.~S. On the problem of hidden variables in quantum mechanics. {\it Rev. Mod. Phys.} {\bf 38,} 447-352 (1966).

\bibitem{Kochen1967}
Kochen, S. \& Specker, E. P. The problem of hidden variables in quantum mechanics. {\it J. Math. Mech.} {\bf 17,} 59-87 (1967).

\bibitem{Klyachko2008}
Klyachko, A.~A., Can, M.~A., Binicio\u{g}lu, S. \& Shumovsky, A.~S. Simple test for hidden variables in spin-1 systems. {\it Phys. Rev. Lett.}  {\bf 101,} 020403 (2008).

\bibitem{Cabello2008}
Cabello, A. Experimentally testable state independent quantum contextuality.  {\it Phys. Rev. Lett.}  {\bf 101,} 210401 (2008).

\bibitem{Araujo2013}
Ara\'{u}jo, M., T\'{u}lio, M., Costantino, Q., Cunha, B.~M.~T. \& Cabello, A. All noncontextuality inequalities for the n-cycle scenario. {\it Phys. Rev. A} {\bf 88,} 022118 (2013).

\bibitem{Yu2012} 
Yu, S. \& Oh, C.H. State-independent proof of Kochen-Specker theorem with 13 rays. {\it Phys. Rev. Lett.} {\bf 108,} 030402 (2012).

\bibitem{Bartosik2009} 
Bartosik, H. \emph{et al.} Experimental test of quantum contextuality in neutron interferometry. {\it Phys. Rev. Lett.} {\bf 103,} 040403 (2009).

\bibitem{Kirchmair2009}
Kirchmair, G. \emph{et al.} State-independent experimental test of quantum contextuality. {\it Nature} {\bf 460,} 494-497 (2009).
	
\bibitem{Zhang2013} 
Zhang, X. M. \emph{et al.} State independent experimental test of quantum contextuality with a single trapped ion. {\it Phys. Rev. Lett.} {\bf 110,} 070401 (2013).
	
\bibitem{Lapkiewicz2011} 
Lapkiewicz, R. \emph{et al.} Experimental non-classicality of an indivisible quantum system. {\it Nature} {\bf 474,} 490-493 (2011).
	
\bibitem{Marques2014b} 
Marques, B., Ahrens, J., Nawareg, M., Cabello, A. \& Bourennane, M. Experimental observation of Hardy-like quantum contextuality. {\it Phys. Rev. Lett.} {\bf 113,} 250403 (2014).
	
\bibitem{Mazurek2015} 
Mazurek, M. D., Pusey, M. F., Kunjwal, R., Resch, K. J. \& Spekkens, R. W. An experimental test of noncontextuality without unwarranted idealizations. {\it Nat. Commun.} {\bf 7,} 11780 (2016).
	
\bibitem{George2013} 
George, R. E. \emph{et al.} Opening up three quantum boxes causes classically undetectable wavefunction collapse. {\it PNAS} {\bf 110,} 3777-3781 (2013).

\bibitem{Kong2016}
Kong, X. M. \emph{et al.} Experimental test of nonclassicality of quantum mechanics using an individual atomic solid-state quantum system. Preprint at http://arxiv.org/abs/1602.02455 (2016).
	
\bibitem{Peres2002} 
Peres, A. {\it Quantum Theory: Concepts and Methods.} Fundamental Theories of Physics, Vol. {\bf 57} (Springer Netherlands, 2002).
	
\bibitem{Arias2015} 
Arias, M. \emph{et al.} Testing noncontextuality inequalities that are building blocks of quantum correlations. {\it Phys. Rev. A} {\bf 92,} 032126 (2015).
	
\bibitem{Larsson2014} 
Larsson, J.-\AA{}. Loopholes in Bell inequality tests of local realism. {\it J. Phys. A: Math. Theor.} {\bf 47}, 424003 (2014).
	
\bibitem{Guhne2010} 
O. G\"{u}hne \emph{et al.} Compatibility and noncontextuality for sequential measurements. {\it Phys. Rev. A} {\bf 81,} 022121 (2010).
	
\bibitem{Szangolies2013} 
Szangolies, J., Kleinmann, M. \& G\"{u}hne, O. Tests against noncontextual models with measurement disturbances. {\it Phys. Rev. A} {\bf 87}, 050101 (2013).
	
\bibitem{Howard2014} 
Howard, M., Wallman, J., Veitch,  V. \& Emerson, J. Contextuality supplies the `magic' for quantum computation. {\it Nature} {\bf 510,} 351-355 (2014).
		
\bibitem{Koch2007}
Koch, J. \emph{et al.} Charge-insensitive qubit design derived from the Cooper pair box. {\it Phys. Rev. A} {\bf 76,} 042319 (2007).
		
\bibitem{Eichler2014}
Eichler, C., Salathe,  Y., Mlynek,  J., Schmidt,  S. \&  Wallraff,  A. Quantum-limited amplification and entanglement in coupled nonlinear resonators. {\it Phys. Rev. Lett.} {\bf 113,} 110502 (2014).

\bibitem{Bianchetti2010}
Bianchetti, R. S. \emph{et al.} Control and tomography of a three level superconducting artificial atom. {\it Phys. Rev. Lett.} {\bf 105,} 223601 (2010).

\bibitem{Jerger2015}
Jerger, M., Macha, P., Hamann, J. R., Reshitnyk,  Y., Juliusson,  K. \& Fedorov A. Realization of a binary-outcome projection measurement of a three-level superconducting quantum system. {\it Phys. Rev. Applied} {\bf 6,} 014014 (2016).

\bibitem{Elkouss2015}
Elkouss, D. \& Wehner, S. Nearly optimal P-values for all Bell inequalities. Preprint at http://arxiv.org/abs/1510.07233 (2015).

\bibitem{Kleinmann2011}
Kleinmann, M., Ghne, O., Portillo, J. R., Larsson,  J. \& Cabello, A. Memory cost of quantum contextuality. {\it New Journal of Physics} {\bf 13,} 113011 (2011).

\bibitem{Meyer1999}
Meyer, D. A. Finite precision measurement nullifies the Kochen-Specker theorem. {\it Phys. Rev. Lett.} {\bf 83,} 3751-3754 (1999).

\bibitem{Kent1999}
Kent, A. Noncontextual hidden variables and physical measurements. {\it Phys. Rev. Lett.} {\bf 83,} 3755-3757 (1999).

\bibitem{Simon2001}
Simon, C., Brukner, \u{C}. \& Zeilinger, A. Hidden-variable theorems for real experiments. {\it Phys. Rev. Lett.} {\bf 86,} 4427-4430 (2001).

\bibitem{Cabello2011}
Cabello, A. \& Terra Cunha, M. Proposal of a two-qutrit contextuality test free of the finite precision and compatibility loopholes. {\it Phys. Rev. Lett.} {\bf 106,}
190401 (2011).

\bibitem{Winter2014}
Winter, A. What does an experimental test of quantum contextuality prove or disprove? {\it J. Phys. A:Math. Theor.} {\bf 47,} 424031 (2014).

\bibitem{Paik2010}
Paik, H.  \emph{et al.} Observation of High Coherence in Josephson Junction Qubits Measured in a Three-Dimensional Circuit QED Architecture.  {\it Phys. Rev. Lett.} {\bf 107,} 240501 (2011).

\end{thebibliography}
\end{document}


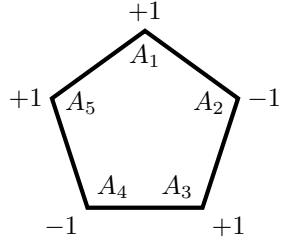
\begin{figure}[t!]
	\begin{center}
		\begin{tikzpicture}[ scale=1.3][H!]
		\draw[ultra thick] (0,1)--(-0.9510565163,0.309017)--(-0.58778525229,-0.809017)--(0.58778525229,-0.809017)--(0.9510565163,0.309017)--cycle;
		\node [above] at (0,1) {$+1$};
		\node [below] at (0,0.95) {$A_1$};
		\node [right] at (0.9510565163,0.309017) {$-1$};
		\node [left] at (0.910565163,0.259017) {$A_2$};
		\node [below right] at (0.58778525229,-0.809017) {$+1$};
		\node [above left] at (0.58778525229,-0.809017) {$A_3$};
		\node [below left] at (-0.58778525229,-0.809017) {$-1$};
		\node [above right] at (-0.58778525229,-0.809017) {$A_4$};
		\node [left] at (-0.9510565163,0.309017) {$+1$};
		\node [right] at (-0.910565163,0.259017) {$A_5$};
		\end{tikzpicture}
	\end{center}
	\caption{The KCBS pentagram, with a specific preassigned set of measurement outcomes that maximize the winning probability}
	\label{fig:pentagon}
\end{figure}

\clearpage

\section*{Supplementary Note 1. The KCBS noncontextuality test as a hypothesis test}
\addtocounter{section}{1}
In this appendix the $P$-value analysis is discussed. First, $P$-values in general and hypothesis tests are briefly discussed. Second, the experiment performed is phrased informally as a game, which will turn out to be convenient for the calculation of a $P$-value. Third, the notation and definitions are written down formally. Fourth, the properties of an $\epsilon$-bounded i.i.d.~noncontextual hidden variable (NCHV) model are explicitly formulated. Fifth, the probability for any $\epsilon$-bounded i.i.d.~NCHV model to win the game is calculated, which is needed for the $P$-value calculation. Sixth, a method to confidently upper bound the amount of incompatibility $\epsilon$ is detailed. Finally, the analysis is directly applied to the experiment performed to obtain a $P$-value.

Tests of any HV model can be phrased as a hypothesis test, in which one aims to reject the null hypothesis that an underlying HV model explains the data. 
The $P$-value is the maximum probability that \emph{if} the null hypothesis were to hold, the data would have been \emph{at least} as extreme as observed. A low $P$-value suggests then that the null hypothesis should be rejected. For the experiment performed, the null hypothesis is that the experiment was governed by an $\epsilon$-bounded i.i.d.~NCHV model (formally defined below), and the $P$-value is the maximum probability that \emph{any} $\epsilon$-bounded i.i.d.~NCHV model could have produced data at least as extreme as observed.

\subsection{KCBS inequality as a win/lose game}
To find an upper bound on the $P$-value, the KCBS inequality is phrased as a win/lose game~\cite{elkouss2015nearly}. The $P$-value can then be expressed as the maximum probability that any $\epsilon$-bounded i.i.d.~NCHV model would have produced \emph{at least} as many wins as observed.
\begin{align}
P\textrm{-value} &= \max_{\mathclap{\substack{\epsilon\textrm{-bounded}\\
\textrm{i.i.d.~NCHV}}}} \Pr\left[{\rm \#\ of\ wins\ at\ least\ as\ large\ as\ observed}\mid \epsilon\textrm{-bounded i.i.d.~NCHV}\right]\ .
\end{align}
If phrased correctly, any $\epsilon$-bounded i.i.d.~NCHV model will have a certain maximum winning probability which is strictly smaller than the maximum winning probability of a contextual model, where in particular $\beta_{\mathrm{win}}$ is the maximum winning probability over all $\epsilon$-bounded i.i.d.~NCHV models~\cite{elkouss2015nearly}. 
\begin{align}
P\textrm{-value} & \leq \sum_{i = c}^n\binom{n}{i}\left(\beta_{\mathrm{win}}\right)^{i}\left(1-\beta_{\mathrm{win}}\right)^{n-i}\label{eq:bound}\ .
\end{align}
The KCBS inequality can be violated by minimizing the correlators of the products of outcomes. This corresponds to a game where a random context is selected, after which the game is won when the two observed outcomes are different.

\subsection{Notation and definitions}
The experiment consists of $n$ trials, where in a specific trial $1\leq l\leq n$ first two ordered measurements are selected. The two measurements have corresponding random variables $\#1^l$ (measured first) and $\#2^l$ (measured second) with outcomes $i$ and $j$, where the possible ordered combinations of $(i,j)$ are restricted to be in the set $\mathcal{D} = \lbrace (1,2), (3,2), (3,4), (5,4), (5,1) \rbrace$. The restriction of the possible ordered contexts to the set $\mathcal{D}$ is due to the fact that the extended KCBS inequality is phrased entirely with these ordered contexts.
Secondly, the outcomes of the two measurements are recorded as $a_i^l, a_j^l$ with corresponding random variables $A_i^l$ and $A_j^l$. The score for a trial is $c^l = \delta(a_i^l+a_j^l) \in \lbrace 0, 1 \rbrace$, i.e.~for the win/lose game described a trial is won when $c^l = 1$, or equivalently, $a_i^l \neq a_j^l$. The total amount of wins $c$ then equals $c = \sum_{l=1}^nc^l$. The random variable $H^l$ includes all previous instances of hidden variables $h$ and the history of all previous in- and outputs $\left(i^l, j^l, a_i^l, a_j^l\right)_{k=1}^{l-1}$.

\subsection{Formulating the null hypothesis}
The null hypothesis that we are testing here is that 
the experiment is governed by an $\epsilon$-bounded i.i.d.~non-contextual hidden variable model, which we will define below.
A hidden-variable model by itself means that, if $\Pr(A_i^l = a_i^l, A_j^l = a_j^l\mid \#1=i, \#2 =j)$ is the probability of observing
outcomes $a_i^l$ and $a_j^l$ during the $l$-th trial of the experiment in which we first measure $i$ and then $j$, then this probability can
be expressed as
\begin{align}
\Pr(A_i^l = a_i^l, A_j^l = a_j^l\mid \#1=i, \#2 =j) = \int {\rm d}\mu(h) \Pr(A_i^l = a_i^l, A_j^l = a_j^l\mid \#1=i, \#2 =j,h)\ ,
\end{align}
where ${\rm d}\mu$ denotes some probability measure over hidden variables $h$.
\\
\noindent Typically, one would make the following assumptions.\\

\noindent 1(a). \textbf{\emph{Uniform randomness}}. Conditioned on a given value $h$ of the hidden variable, the probability of selecting an ordered context $(i,j)\in\mathcal{D}$ in each trial $l$ is uniform
\begin{gather}
\forall~(i,j)\in \mathcal{D},\forall~l,~\Pr(\#1^l=i^l,~\#1^l=j^l,h) = \frac{1}{5}\ .
\end{gather}
In the actual performed experiment the measurements were measured in a predetermined fashion. The uniform randomness criteria must then be augmented by\\

\noindent 1(b). \textbf{\emph{Independent and identically distributed (i.i.d)}}. 
For any context $(i,j)$ 
the outcomes of the experiment during each trial are independent of the history of the experiment
\begin{gather}
\forall(i, j, l), ~A_i^l\indep H^l,~A_j^l\indep H^l \ ,
\end{gather}
and have equal probability distributions for all $l$. 
This is equivalent to saying that the distribution over hidden variables determining the outcomes is the same in every trial. \\

\noindent 2. \textbf{\emph{Sequentiality of the measurements}}. Each of the measurements are separated and ordered in time, so that causality prevents signalling from any of the future attempts to current or previous attempts. In particular, this also implies that if $h$ is the hidden variable of the present trial
\begin{gather}
\Pr(A_i=a_i, A_{j}=a_j|\#1 = i, \#2 = j,h) = \Pr(A_i=a_i|\#1=i,h)\Pr(A_j=a_j|\#1=i,\#2=j,h) \ ,
\end{gather}
since the outcome of the first measurement cannot depend on whether or not a measurement will be performed in the future.\\

\noindent 3. \textbf{\emph{Bounded incompatibility/noncontextuality}}. 
The incompatibility is $\epsilon$-bounded~\cite{guhne2010compatibility}, in the sense that for any instance $h$ of the hidden variable 
and any context $(i,j)$, there exists $\epsilon_{(i,j),h}$ such that\footnote{Note the lack of a factor of a half which is standard in the literature.}
\begin{gather}
-\epsilon_{(i,j),h} \leq \Pr(A_j=a_j|\#1=j,h)-\Pr(A_j=a_j|\#1=i,\#2=j,h) \leq \epsilon_{(i,j),h}\ ,
\end{gather}
with $\epsilon_{(i,j),h} \in \lbrace 0,1\rbrace$, and hence
\begin{gather}
-\epsilon_{(i,j)}\leq \Pr(A_j=a_j|\#1=j)-\Pr(A_j=a_j|\#1=i,\#2=j) \leq \epsilon_{(i,j)}\ ,
\end{gather}
for $\epsilon_{(i,j)} := \int\mathrm{d}\mu(h)\epsilon_{(i,j),h}$ being the average taken over the hidden variables 
such that the $\emph{average incompatibility}$ is $\epsilon$-bounded,
\begin{gather}
-\epsilon \leq \frac{1}{5}\int\mathrm{d}\mu(h) \sum_{(i,j)\in \mathcal{D}} \epsilon_{(i,j),h} = \frac{1}{5}\sum_{(i,j)\in \mathcal{D}} \epsilon_{(i,j)} \leq \epsilon \ \label{eq:epsbound}.
\end{gather}
What we will test here is the null hypothesis of a hidden variable model satisfying (1a), (1b), (2) and (3), which we will call an $\epsilon$-bounded i.i.d.~NCHV model. We note that the analysis of~\cite{elkouss2015nearly} allows arbitrary memory for the hidden-variable model, that is, the i.i.d.~assumption is not needed. However, in the present experiment this assumption is necessary.

\subsection{Upper bounding $\beta_{\mathrm{win}}$}
In this section the maximum winning probability $\beta_{\mathrm{win}}$ is upper bounded for an $\epsilon$-bounded i.i.d.~NCHV model as specified above. 
\begin{align}
\beta_{\mathrm{win}} &= \int\mathrm{d}\mu(h)\sum_{(i,j) \in \mathcal{D}}\Pr(\#1 = i, \#2 = j,h)\Pr(\mathrm{win}|\#1 = i, \#2 = j,h)\label{eq:apstep1}\\
&= \frac{1}{5}\int\mathrm{d}\mu(h)\sum_{(i,j) \in \mathcal{D}}\Pr(\mathrm{win}|\#1 = i, \#2 = j)\label{eq:apstep2}\\
&= \frac{1}{5}\int\mathrm{d}\mu(h)\sum_{(i,j)\in\mathcal{D}}\sum_{a\in\lbrace-1,1 \rbrace}\Pr(A_i=a, A_{j}=-a|\#1 = i, \#2 = j,h)\label{eq:apstep3}\\
&= \frac{1}{5}\int\mathrm{d}\mu(h)\sum_{(i,j)\in\mathcal{D}}\sum_{a\in\lbrace-1,1 \rbrace}\Pr(A_i=a|\#1=i,h)\Pr(A_j=-a|\#1=i,\#2=j,h) \label{eq:apstep4}\\
&= \frac{1}{5}\int\mathrm{d}\mu(h)\sum_{(i,j)\in\mathcal{D}}\Pr(A_i=1|\#1=i,h)\Pr(A_j=-1|\#1=i,\#2=j,h) \nonumber\\
& \hspace{27.1mm}+\Pr(A_i=-1|\#1=i,h)\Pr(A_j=1|\#1=i,\#2=j,h)\label{eq:apstep5}\\
&= \frac{1}{5}\int\mathrm{d}\mu(h)\sum_{(i,j)\in\mathcal{D}}\Pr(A_i=1|\#1=i,h)\left(1-\Pr(A_j=1|\#1=i,\#2=j,h)\right) \nonumber\\
& \hspace{27.1mm}+\left(1-\Pr(A_i=1|\#1=i,h)\right)\Pr(A_j=1|\#1=i,\#2=j,h)\label{eq:apstep6}\\
&= \frac{1}{5}\int\mathrm{d}\mu(h)\sum_{(i,j)\in\mathcal{D}}\Pr(A_i=1|\#1=i,h)\nonumber \\
& \hspace{27.1mm}+\left(1-2\Pr(A_i=1|\#1=i,h)\right)\Pr(A_j=1|\#1=i,\#2=j,h)\label{eq:apstep7} \\
& \leq \frac{1}{5}\int\mathrm{d}\mu(h)\sum_{(i,j)\in\mathcal{D}}\Pr(A_i=1|\#1=i,h)\nonumber\\
&\hspace{27.1mm}+\left(1-2\Pr(A_i=1|\#1=i,h)\right)\Pr(A_j=1|\#1=j,h)+\epsilon_{(i,j),h}\label{eq:apstep81}\\
& \leq \epsilon + \frac{1}{5}\int\mathrm{d}\mu(h)\sum_{(i,j)\in\mathcal{D}}\Pr(A_i=1|\#1=i,h)\nonumber\\
&\hspace{27.1mm}+\left(1-2\Pr(A_i=1|\#1=i,h)\right)\Pr(A_j=1|\#1=j,h)\label{eq:apstep91}
\end{align}
where in equation \eqref{eq:apstep1} the winning probability is written as an integral over hidden variables $h$, and a conditional sum over selecting measurements $i$ and $j$. Equation \eqref{eq:apstep2} follows from $\Pr(\#1 = i, \#2 = j|h) = \frac{1}{5},~\forall (i, j)\in \mathcal{D}, \forall h$. In the case of the present experiment, we rely on condition (1b). That is, there is no random choice \emph{a priori}, but instead such random selection is simulated after the fact.

In equation \eqref{eq:apstep3} the winning condition is formulated in the summands for a fixed $a \in \lbrace -1,1\rbrace$. Equation \eqref{eq:apstep4} is based on the fact that the outcome of $A_i$ cannot depend on $\#2$ since it hasn't been measured yet. In equation \eqref{eq:apstep5} the terms depending on $a\in\lbrace-1,1\rbrace$ are written out explicitly. Equations \eqref{eq:apstep6} and \eqref{eq:apstep7} follow from $\Pr(A_i=1|\#1=i) = \left(1-\Pr(A_i=-1|\#1=i)\right)$ and rewriting. The inequalities in equations \eqref{eq:apstep81} and \eqref{eq:apstep91} follow from the inequalities 
\begin{gather}
-\epsilon_{(i,j),h} \leq \Pr(A_j=a_j|\#1=j,h)-\Pr(A_j=a_j|\#1=i,\#2=j,h) \leq \epsilon_{(i,j),h}\ ,  \\
 \frac{1}{5}\int\mathrm{d}\mu(h) \sum_{(i,j)\in \mathcal{D}} \epsilon_{(i,j),h} \leq \epsilon \ .
\end{gather}
Note that the values $\epsilon_{(i,j),h}$ cannot be known, but fortunately $\epsilon$, the parameter of relevance, \emph{can} be bounded from above as will be shown in the next section. 
The integral term equals
\begin{align}
&\hspace{4.3mm}\int\mathrm{d}\mu(h)\sum_{(i,j)\in\mathcal{D}}\Pr(A_i=1|\#1=i,h)+\left(1-2\Pr(A_i=1|\#1=i,h)\right)\Pr(A_j=1|\#1=j,h)\\
& = \int\mathrm{d}\mu(h)\sum_{(i,j)\in\mathcal{D}}\Pr(A_i=1|\#1=i,h)+\Pr(A_j=1|\#1=j,h)\nonumber \\
&\hspace{29mm}-2\Pr(A_i=1|\#1=i)\Pr(A_j=1|\#1=j,h)\label{eq:apstep9}\\
& = 2\int\mathrm{d}\mu(h)\left(\sum_{i=1}^5\Pr(A_i=1|\#1=i,h)-\sum_{(i,j)\in\mathcal{D}}\Pr(A_i=1|\#1=i,h)\Pr(A_j=1|\#1=j,h)\right)\label{eq:apstep10}\\
& \leq 4 \label{eq:apstep12}\ ,
\end{align}
where equation \eqref{eq:apstep9} follows from rewriting and equation \eqref{eq:apstep10} follows from the fact that 
\begin{gather}
\sum_{(i,j)\in\mathcal{D}} \Pr(A_i=1|\#1=i,h) +\Pr(A_j=1|\#1=j,h) = 2\sum_{i = 1}^5\Pr(A_i=1|\#1=i,h)\ .
\end{gather}
The integral term achieves its maximum when all the probability mass $\mathrm{d}\mu(h)$ is concentrated on the deterministic distributions (i.e.~$\forall i,~\Pr(A_i=1|\#1=i,h) \in \lbrace0,1 \rbrace$) that maximize equation $\eqref{eq:apstep10}$. One can easily see that the best strategy is to alternate the outcomes as in Figure \ref{fig:pentagon}. For these distributions, the sum achieves its maximum value of $2$, from which the upper bound in equation \eqref{eq:apstep12} follows. Combining equations \eqref{eq:apstep12} and \eqref{eq:apstep91} yields
\begin{gather}
\beta_{\mathrm{win}} \leq \frac{4}{5} + \epsilon\ .
\end{gather}

\subsection{Upper bounding the incompatibility $\epsilon$}
If the assumption is made that the experiment behaves in an i.i.d fashion, then the average incompatibility $\epsilon$ is a parameter that can be estimated by performing a game separate from the main experiment. Informally, the game is played over $n$ trials, where a total score $g_{\textrm{avg}}^n$ is assigned for the whole game. By construction, the absolute value of the expectation value of the score (i.e.~$\left|\mathds{E}\lbrack G_{\textrm{avg}}^n\rbrack\right|$) will be an estimate for $\epsilon$.

However, since there are only a finite amount of trials the observed score will deviate from its expectation value, which is the quantity of interest. As will be shown, the maximum deviation $t$ between the observed score $g_{\textrm{avg}}^n$ and its expectation value $\mathds{E}\lbrack G_{\textrm{avg}}^n\rbrack$ can be bounded with high probability. This allows for the $\epsilon$ parameter to be upper bounded with high probability by the absolute value of the observed score (i.e.~$\left|g_{\textrm{avg}}^n\right|$), plus some error margin $t$.\\

\noindent More formally, a game is played consisting of $n$ trials, where in each trial $l$ an ordered context $(i,j) \in \mathcal{D}$ is uniformly selected. Then either $(i,j)$ or $(j,i)$ is measured, depending on the outcome of a uniform random variable $X^l$ taking values $x^l = 1$ or $x^l = -1$, respectively. The outcome $a_j^l$ corresponding to the random variable $A_j^l$ is recorded. The concrete score $g^l$, corresponding to the random variable $G^l$, is equal to $2$ when $x^l = 1,~a_j^l =1$, equal to $-2$ when $x^l = -1, ~a_j^l = 1$ and $0$ for all other instances. The average score at the end of $n$ trials equals
\begin{gather}
g_{\textrm{avg}}^n = \frac{1}{n}\sum_{l=1}^{n}g^l
\end{gather}
with the associated random variable $G_{\textrm{avg}}^n$. The expectation value of $G_{\textrm{avg}}^n$ taken over all possible contexts $(i,j)\in \mathcal{D}$ and values of $X^l$ satisfies
\begin{gather}
\mathds{E}\left[G_{\textrm{avg}}^n\right] = \frac{1}{5}\sum_{(i,j)\in \mathcal{D}}\left(\Pr(A_j=1|\#1=j)-\Pr(A_j=1|\#1=i,\#2=j)\right)\ \label{eq:expG}.
\end{gather}
From equations \eqref{eq:epsbound} and \eqref{eq:expG} $\epsilon$ can be estimated by $\left|\mathds{E}\lbrack G_{\textrm{avg}}^n\rbrack\right|$, i.e.~the absolute value of the expectation value of the score over $n$ trials. Since there is only a finite amount of samples $\mathds{E}\lbrack G_{\textrm{avg}}^n\rbrack$ cannot be estimated perfectly, so that for any experiment the observed score $g_{\textrm{avg}}^n$ will deviate from the expectation value $\mathds{E}\lbrack G_{\textrm{avg}}^n\rbrack$. That is, $g_{\textrm{avg}}^n-t \leq \mathds{E}\lbrack G_{\textrm{avg}}^n\rbrack \leq g_{\textrm{avg}}^n+t$, or equivalently, $\left|
g_{\textrm{avg}}^n-\mathds{E}\lbrack G_{\textrm{avg}}^n\rbrack\right|\leq t$ for some $t>0$. The probability that the observed average value corresponding to a sequence of i.i.d.~random variables (the average score $g_{\textrm{avg}}^n$ in the present case) deviates at most $t$ from the expectation value of the average of those random variables ($\mathds{E}\lbrack G_{\textrm{avg}}^n\rbrack$ in the present case) can be upper bounded with Bentkus' inequality~\cite{bentkus2004hoeffding}.
Before stating Bentkus' inequality, we define
\begin{align}
P_{n,k}\left(\mathbb{B}_{\gamma}\right) := \sum_{i=k}^{n}\binom{n}{i}\gamma^i\left(1-\gamma\right)^{n-i}\ 
\end{align}
and
\begin{align}
\mathring{P}_{n,y}\left(\mathbb{B}_{\gamma}\right) = \left(P_{n,\floor*{y}}\left(\mathbb{B}_{\gamma}\right)\right)^{1-(y-\floor*{y})}\left(P_{n,\ceil*{y}}\left(\mathbb{B}_{\gamma}\right)\right)^{y-\floor*{y}}\ .
\end{align}\\

\begin{theorem}[Bentkus' inequality] Let $M^1,M^2,\ldots,M^n$ be a martingale sequence with differences $X^l = M^l - M^{l-1}$ and $M^0 = 0$. If for $l = 1\ldots n$ the differences satisfy the following boundedness condition,
\begin{align}
\Pr(-\alpha^l\leq X^l\leq 1-\alpha^l) = 1\ ,
\end{align}
then
\begin{align}
\Pr(M^n\geq t)\leq e\mathring{P}_{n,t+n\gamma}\left(\mathbb{B}_{\gamma}\right)
\end{align}
with $\gamma = \sum_{i=1}^n\alpha_i/n$.
\end{theorem}
\noindent As we show in the corollary below, Bentkus' inequality can be phrased in a way to get a bound on $\Pr\left(\left|g_{\textrm{avg}}^n-\mathds{E}\lbrack G_{\textrm{avg}}^n\rbrack\right|\geq t\right)$.
\begin{corollary}
Let $X_1,X_2,\ldots,X_n$ be a sequence of i.i.d.~variables satisfying $-a\leq X_i\leq a,~\forall i\in \lbrace 1,2,\ldots, n\rbrace$. Define $\overline{X} = \frac{1}{n}\sum_{i=1}^nX^i$, then
\begin{gather}
\Pr\left(\left|\overline{X}-\mathds{E}\lbrack \overline{X}\rbrack\right|\geq t\right) \leq 2e\mathring{P}_{n,n\left(\frac{t+2a}{4a}\right)}\left(\mathbb{B}_{1/2}\right)\label{eq:Bentkus2}\ .
\end{gather}
\end{corollary}
\begin{proof}
Define the sequence $M^1,M^2,\ldots,M^n$ with $M^l = \frac{1}{4a}\sum_{i=1}^l\left(X^i-\mathds{E}\left[\overline{X}\right]\right)$. This is a martingale sequence with $\mathds{E}\lbrack M^{l+1}|M^1,M^2,\ldots,M^l\rbrack = 0$ since the sequence $X^1,X^2,\ldots,X^n$ is i.i.d., and has bounded differences $-\frac{1}{2} \leq M^l-M^{l-1} = \frac{1}{4a}\left(X^l-\mathds{E}\lbrack \overline{X}\rbrack\right)\leq \frac{1}{2}$ since $-a\leq X^l\leq a$. Bentkus' inequality can then be applied to the sequence $M^1,M^2,\ldots,M^n$ with $\gamma = \sum_{i=1}^n\alpha/n = 1/2$,
\begin{gather}
\Pr\left(M^n\geq \frac{nt}{4a}\right) \leq e\mathring{P}_{n,\frac{nt}{4a}+\frac{n}{2}}\left(\mathbb{B}_{1/2}\right) = e\mathring{P}_{n,n\left(\frac{t+2a}{4a}\right)}\left(\mathbb{B}_{1/2}\right)\ .
\end{gather}
Rewriting yields
\begin{align}
\Pr\left(M^n\geq \frac{nt}{4a}\right) &= \Pr\left(\frac{1}{4a}\sum_{i=1}^n\left(X_i-\mathds{E}\lbrack \overline{X}\rbrack\right)\geq \frac{nt}{4a}\right)\\
&= \Pr\left(\sum_{i=1}^n\left(X_i-\mathds{E}\lbrack \overline{X}\rbrack\right)\geq nt\right)\\
&= \Pr\left(n\left(\overline{X}-\mathds{E}\lbrack \overline{X}\rbrack\right)\geq nt\right)\\
&= \Pr\left(\overline{X}-\mathds{E}\lbrack \overline{X}\rbrack\geq t\right)\ ,
\end{align}
so that
\begin{gather}
\Pr\left(\overline{X}-\mathds{E}\lbrack \overline{X}\rbrack\geq t\right) \leq e\mathring{P}_{n,n\left(\frac{t+2a}{4a}\right)}\left(\mathbb{B}_{1/2}\right)\ .
\end{gather}
The same procedure can be followed for the martingale sequence $-M^1,-M^2,\ldots,-M^n$, so that
\begin{gather}
\Pr\left(\left|\overline{X}-\mathds{E}\lbrack \overline{X}\rbrack\right|\geq t\right) \leq 2e\mathring{P}_{n,n\left(\frac{t+2a}{4a}\right)}\left(\mathbb{B}_{1/2}\right)\ .
\end{gather}
\end{proof}
\noindent In other words, with equation \eqref{eq:Bentkus2} it is possible to obtain a \emph{confidence interval}, relating the maximum deviation one would expect to see between the average of i.i.d.~variables and their expectation value for a certain probability. In particular, it can be used to bound the probability that $\left|g_{\textrm{avg}}^n-\mathds{E}\lbrack G_{\textrm{avg}}^n\rbrack\right| \geq t$. Specifically, we apply equation \eqref{eq:Bentkus2} to the sequence of i.i.d~random variables $G^1,G^2,\ldots,G^n$, where the mean over the random variables is $\frac{1}{n}\sum_{l=1}^{n}G^l = G_{\textrm{avg}}^n$, and $-2 \leq G^l \leq 2$ so that $a = 2$. Equation \eqref{eq:Bentkus2} then yields
\begin{gather}
\Pr\left(\left|G_{\textrm{avg}}^n-\mathds{E}\lbrack G_{\textrm{avg}}^n\rbrack\right|\geq t\right) \leq 2e\mathring{P}_{n,n\left(\frac{t+4}{8}\right)}\left(\mathbb{B}_{1/2}\right)\label{eq:Bentkus3}\ .
\end{gather}
In particular, equation \eqref{eq:Bentkus3} can be applied to the observed score $g_{\textrm{avg}}^n$, which allows for a parameter estimation of $\epsilon$ with some confidence interval $t$, 
\begin{gather}
\Pr\left(\left|g_{\textrm{avg}}^n-\mathds{E}\lbrack G_{\textrm{avg}}^n\rbrack\right|\geq t\right) \leq 2e\mathring{P}_{n,n\left(\frac{t+4}{8}\right)}\left(\mathbb{B}_{1/2}\right)
\end{gather}
That is, with probability less than $2e\mathring{P}_{n,n\left(\frac{t+4}{8}\right)}\left(\mathbb{B}_{1/2}\right)$, the observed data $g_{\textrm{avg}}^n$ will satisfy $\left|g_{\textrm{avg}}^n-\mathds{E}\lbrack G_{\textrm{avg}}^n\rbrack\right|\geq t$. Using equations \eqref{eq:epsbound} and \eqref{eq:expG}, $\epsilon$ can then be upper bounded by $\left|g_{\textrm{avg}}^n\right|+t$ with probability greater than $1-2e\mathring{P}_{n,n\left(\frac{t+4}{8}\right)}(\mathbb{B}_{1/2})$.
We note that we are free to choose $t$ and $n$ independent of recorded data to aim for a bound on the probability that is good enough to confidently upper bound $\epsilon$ by $\left|g_{\textrm{avg}}^n\right|+t$. 
Setting $t = 0.005$ with $n = 9207101$ measurements (which constitutes the first half of the data collected), we find

\begin{align}
\Pr\left(\left|g_{\textrm{avg}}^n-\mathds{E}\lbrack G_{\textrm{avg}}^n\rbrack\right|\geq t\right) &\leq 2e\mathring{P}_{n,n\left(\frac{t+4}{8}\right)}\left(\mathbb{B}_{1/2}\right)\\
& \leq 4.1\cdot 10^{-4}\ .
\end{align}
\newpage
\subsection{Statistical analysis}
The analysis above can then directly be applied to the performed contextuality experiment. The total analysis consists of the estimation of $\epsilon$ and the testing of the null hypothesis as formulated above. These two tests are performed on two disjoint parts of the data. Recall that we take it as a given that the NCHV model is i.i.d., that is, the devices perform the  same in every single trial. This allows us to make separated estimates. 

Before proceeding, we clarify that the contexts were not chosen randomly during the course of the experiment, but note that making random measurements is not necessary when testing an i.i.d.~model.
Furthermore, the $P$-value analysis of~\cite{elkouss2015nearly} we use here assumes that the number of trials $n$ is selected independently of the data.
Here, we had to make an explicit assumption that we were able to pick $n$ uninfluenced by an HV model, since the data had already been taken.
We again note that the model to be tested is i.i.d., and also that it has no memory. 
We emphasize that no such assumptions were made in the analysis of the recent loophole-free Bell test~\cite{hensen2015loophole}, where in particular the model was allowed to have full memory.

As shown in the previous section, with probability less than $4.1\cdot 10^{-4}$ the condition $\epsilon\leq \left|g_{\textrm{avg}}^n\right| + 0.005$ will hold.
Calculating $g_{\textrm{avg}}^n$ from recorded data we find $\left|g_{\textrm{avg}}^n\right|\leq 0.036286$,
so that $\beta_{\mathrm{win}} \leq \frac{4}{5} + \left|g_{\textrm{avg}}^n\right| + t = \frac{4}{5} + 0.036286 + 0.005 = 0.841286$.

Second, we can now compute the $P$-value of testing an $\epsilon$-bounded i.i.d.~NCHV. 
We remark that for large $n$ as in the present experiment, the distribution is approximately normal, meaning that an estimate of the $P$-value based on standard deviations is approximately accurate. However, the analysis of~\cite{elkouss2015nearly} gives a \emph{tight} bound on the $P$-value.
This can be calculated using the results of~\cite{elkouss2015nearly} by setting $\beta_{\mathrm{win}} = \frac{4}{5} + \left|g_{\textrm{avg}}^n\right|+t = 0.841286$, counting the amount of wins $c$ and trials $n$ and applying equation \eqref{eq:bound}. 
We have $c = 3912769$, and $n = 4603450$, yielding
\begin{align}
P\textrm{-value} & \leq \sum_{i = c}^n\binom{n}{i}\left(\beta_{\mathrm{win}}\right)^{i}\left(1-\beta_{\mathrm{win}}\right)^{n-i}\\
& \leq 2.96\cdot 10^{-575}\ .
\end{align}